\documentclass[preprint]{elsarticle}

\usepackage{lineno}
\usepackage{hyperref}
\usepackage{amsfonts}
\usepackage{amsmath}
\usepackage{amssymb}
\usepackage{amsthm}
\usepackage{graphicx} 
\usepackage{hhline}
\usepackage{wrapfig}

\usepackage{subfigure} 
\usepackage{subfigmat} 
\usepackage{hyperref} 
\usepackage{breqn} 
\usepackage{multirow}
\usepackage[inline]{enumitem}
\usepackage{siunitx}
\usepackage{dcolumn}           
\usepackage{secdot}
\usepackage[ruled,vlined]{algorithm2e}
\usepackage{float}
\usepackage{csquotes}
\usepackage{bm}
\usepackage{mathtools}
\DeclarePairedDelimiter\abs{\lvert}{\rvert}%
\DeclarePairedDelimiter\norm{\lVert}{\rVert}%

\makeatletter
\let\oldabs\abs
\def\abs{\@ifstar{\oldabs}{\oldabs*}}
\let\oldnorm\norm
\def\norm{\@ifstar{\oldnorm}{\oldnorm*}}
\makeatother

\makeatletter
\newcommand{\apxsub}[3]{%
\subsection*{#1 #2}
\addcontentsline{toc}{subsection}{#1 #2}
\def\@currentlabel{#1}%
\label{#3}%
}
\makeatother

\journal{Acta Astronautica}

\begin{document}

\begin{frontmatter}
\title{Autonomous orbit determination for satellite formations using relative sensing: observability analysis and optimization\tnoteref{mytitlenote}}%

\author[mymainaddress]{Pedro Rocha Cachim\corref{mycorrespondingauthor}} \fnref{myfootnote}
\ead{pedrocachim@tecnico.ulisboa.pt}
\author[mymainaddress]{João Gomes\corref{mycorrespondingauthortwo}}
\ead{jpg@isr.tecnico.ulisboa.pt}
\author[mymainaddress]{Rodrigo Ventura\corref{mycorrespondingauthorthird}}
\ead{rodrigo.ventura@isr.tecnico.ulisboa.pt}
\address[mymainaddress]{Institute for Systems and Robotics, Instituto Superior Técnico, Av. Rovisco Pais 1,
1049-001 Lisbon, Portugal}
\fntext[myfootnote]{Corresponding author.}





\begin{abstract}
Orbit determination of spacecraft in orbit has been mostly dependent on either GNSS satellite signals or ground station telemetry. 
Both methods present their limitations, however: GNSS signals can only be  used effectively in earth orbit, and ground-based orbit determination presents an inherent latency that increases with the Earth-spacecraft distance.
For spacecraft flying formations, an alternative method of orbit determination, independent of external signals, consists in the observation of the spacecraft's position with respect to the central body through the relative positioning history of the spacecraft within the formation.
In this paper, the potential of the relative positioning method is demonstrated in the context of the SunRISE mission, and compared with the mission's previously proposed orbit determination methods. An optimization study is then made to find the optimal placement of a new spacecraft in the formation so as to maximize the positioning accuracy of the system. Finally, the possibility of removing part of the system's relative bearing measurements while maintaing its observability is also studied. The resulting system is found to be observable, but ill-conditioned.
\end{abstract}

\begin{keyword}
Spacecraft formation flying \sep relative positioning \sep trajectory optimization \sep observability study
\end{keyword}

\end{frontmatter}


\section{Introduction}

 Ever since the launch of the first space exploration mission, much effort has been directed towards the goal of lowering the cost of these missions. One of the concepts developed with this purpose in mind was that of spacecraft formation flying~\cite{fraser2019adaptive}.
 The fractioning of a single large satellite's payload and operational functions into several smaller elements should not only lead to a reduction in cost, but also to a more reliable mission design, since failure of one spacecraft would not necessarily imply the collapse of the system~\cite{liu2018survey}. Spacecraft FF (Formation Flying) may also allow for otherwise impractical mission concepts with a single spacecraft to become more achievable~\cite{nag2016effect}. This is particularly the case for applications requiring wide and precise baseline separations such as interferometry and gravimetry~\cite{fraser2019adaptive}.

The design of spacecraft FF mission's GNC (guidance, navigation and control) systems is particularly challenging. Most navigation solutions employed by FF missions have relied on GNSS receivers. Even the MMS mission, in which the formation reached orbital apogees as far as 25 Earth radii, ultimately used GNSS receivers tuned for receiving low strength GNSS signals in order to perform orbit determination at higher altitudes~\cite{winternitz2016gps}. 

For deep space missions, the design of orbit determination systems typically relies on ground station telemetry~\cite{montenbruck2002satellite}. However, the inherent latency of ground station signals for such missions does not allow for knowledge of the real-time state. Without the presence of a navigation constellation in orbit of the central body, overcoming this challenge requires the development of autonomous, real-time navigation methods~\cite{dutta2014statistical}. 

One such method proposed for FF missions is based solely on relative positioning. As demonstrated by Markley, knowledge of the time history of the relative positions of spacecraft in orbit in an inertially aligned frame may allow for the determination of the absolute position of the spacecraft with respect to the central body~\cite{markley1987autonomous}.
This method has aroused much interest due to its potential for real-time autonomous navigation, independent of any external telemetry.

In his research, Markley concluded that the system would only be unobservable if the two spacecraft have equal altitude time histories, and are neither coplanar nor oriented such that they cross the line of intersection of the two orbital planes simultaneously. Further studies have been done on the observability of this system.
For example, Psiaki studied the use of this method to autonomously determine the position of the two spacecraft while also estimating gravity model parameters of the celestial body around which the spacecraft orbit~\cite{psiaki2011absolute}. 
The use of this method has also been proposed as a solution to increase the autonomy of GNSS satellite ephemeris error update from ground infrastructure~\citep{yu2019autonomous}.

Within the topic of absolute positioning through relative positioning measurements, the question of which orbital configuration will optimize the accuracy of the system has already been addressed in the literature. Psiaki made observations on how the orbital parameters affect the observability/performance of the system of a two-element system~\cite{psiaki1999autonomous}. Ou also studied the impact of the formation's absolute orbital elements on the observability within a two-element formation \cite{ou2016autonomous}. Subsequent studies by Ou focused on designing an autonomous navigation scheme for Mars exploration with optimized performance indexes based on the observability matrix or in the  Fisher Information Theory~\cite{ou2018observability,ou2018absolute}.
In \cite{li2019obs}, Li contributes to the development of the link between observability analysis and navigation performance by proposing observability-based modified versions of nonlinear estimators applied to this problem.

In order for two spacecraft to directly observe their relative position, range and LOS (Line-Of-Sight) vector  measurements need to be combined.  Studies have also been done on the observability of a system with either ranging-only \cite{hill2007autonomous} or LOS vector-only \cite{yim2004autonomous}. 

According to Yim's work \cite{yim2004autonomous}, relative LOS vector measurements with inertial attitude information allow for the system to be fully observable, even without perturbations from the Earth's oblateness effect on its gravity field. The only unobservable scenarios take place when the two spacecraft are in the same orbiting plane with no inclination. Increasing the complexity of the gravity field of the central body or the overall inclination of the formation may improve the system's observability. More recent studies have further analysed the observability of the system under a variable number of spacecraft, orbital configurations and dynamic models~\cite{hu2020autonomous, hu2021distributed}.

Hill found that ranging measurements in the two-body problem may at best observe the shape, phase and relative orientation of the orbits of the two spacecraft, but not the absolute orientation with respect to the inertial frame of reference, due to the spherical symmetry of the two-body problem's gravity field model \cite{hill2007autonomous}. When more complex and asymmetrical gravity fields are considered, the problem may become observable~\cite{hill2007autonomous, hill2008autonomous}. This system may also be rendered observable when considering three-body problem dynamics, with one of the spacecraft placed in the Lagrange points 1 or 2. This navigation concept for the Earth-Moon system is known as LiAISON~, and has been discussed as an autonomous navigation solution for vehicles on the far side of the lunar surface~\cite{hill2006linked, wang2019liaison}. 

Some research has also been done on the potential of ranging-only autonomous navigation for GNSS satellite ephemeris corrections, concluding that, although the inclusion of perturbation dynamic models such as third-body gravity and solar radiation pressure may allow for full observability of the system, its accuracy is very dependent on the perturbation dynamic model error~\cite{li2020observability}. 


This paper seeks to contribute to the research effort towards the autonomous orbit determination through relative sensing methods. Due to the sparseness of deep-space flying formation concepts where its use would be most beneficial, the near-GEO SunRISE mission concept was chosen as a case study~\citep{alibay2017sun}. 

The following three topics of discussion were set for this study:

\begin{itemize}
	\item Analysis and comparison of a new relative positioning autonomous navigation method with the previously proposed navigation solutions for the SunRISE concept discussed in \citep{stuart2017formation};
	\item Placement of an additional spacecraft in the formation in a configuration that optimizes the autonomous navigation system's performance;
	\item Study of suppression of system sensors while maintaining full state observability.
\end{itemize}

The main contribution of this paper consists in the study on the feasibility of the use of the autonomous relative positioning orbit determination method in the context of the SunRISE mission, and its ability to meet the mission's navigation requirements. 
We consider the addition of a new spacecraft to improve the positioning performance. The optimization of the new spacecraft's orbital configuration  provides insight into how it affects the observability of the autonomous relative positioning orbit determination method.
Finally, the paper also shows that the system in question may be capable of retaining its observability when deprived of most of its relative bearing measurement systems.

In Section \ref{sec:locmet}, the new orbit determination solution for the SunRISE mission is proposed and compared with the solutions previously studied in \citep{stuart2017formation}. The new spacecraft placement optimization study is described in Section \ref{sec:ObsOpt}, followed by the study of the potential elimination of system sensors in Section \ref{sec:systsensred}. The evaluation of the results in a simulation environment is then presented in Section \ref{sec:simres} and the main conclusions are drawn in Section \ref{sec:concl}.

\section{Orbit Determination Methods}
\label{sec:locmet}

The SunRISE mission, around which this study will be focused, is a NASA JPL mission aiming to study Coronal Mass Ejections from the sun, mainly how solar energetic particles are released. To accomplish this goal, the mission was designed as a flying formation with 6 identical 6U CubeSats forming an observatory in a 25-hour, near geostationary orbit. The interspacecraft distances range from $\sim 1-10$ km along an orbit, with passive formation keeping. For the mission's scientific objectives to be met, a maximum of 3 m relative positioning accuracy was defined~\cite{alibay2017sun}.


Two solutions were discussed in \cite{stuart2017formation} for the positioning system of the SunRISE mission: a GNSS-based method, and a RF/vision-based method. Although the study concluded that the GNSS solution would be preferable for the mission, it also recognized the potential for improvement of the RF/Vision-based method. We will therefore briefly describe the discarded RF/Vision-based method, and propose an alternative filtering solution.


\subsection{RF/vision-based method}
\label{subsect:rfvis}

The RF/vision-based method described in \cite{stuart2017formation} proposes to use UHF crosslinks to obtain pseudorange measurements between the spacecraft in the formation, along with a star tracker to measure the orientation of the spacecraft relative to the stars, as well as the relative direction/bearing of another spacecraft with respect to that same inertial stellar frame of reference.

\subsubsection{Crosslink schedule}

Because this method considers the constraint that relative measurements can only be obtained beween pairs of satellites, a measurement schedule is necessary to guarantee that every possible spacecraft pair is covered within a measurement cycle. The cycle proposed in \cite{stuart2017formation} is described in Table \ref{tab:meassched1}, in which the spacecraft are numbered from 1 to 6. For each 1-minute measurement slot, observations are made every second. The 9 minute interval between sets accounts for the necessary time for the spacecraft to slew to point their cameras towards the new target and for the radios of the new pair to lock onto each other.

\begin{table}[H]
\centering
\begin{tabular}{|l|c|}
\hline
\multicolumn{1}{|c|}{\textbf{Time Interval, min}} & \textbf{S/C Pairs} \\ \hline
\multicolumn{1}{|c|}{$[t_0+9+50k, t_0+10+50k]$} & \textbf{\begin{tabular}[c]{@{}c@{}}1-2 \ 3-4 \ 5-6\end{tabular}} \\ \hline
$[t_0+19+50k, t_0+20+50k]$ & \textbf{\begin{tabular}[c]{@{}c@{}}1-3 \ 2-5 \ 4-6\end{tabular}} \\ \hline
$[t_0+29+50k, t_0+30+50k]$ & \textbf{\begin{tabular}[c]{@{}c@{}}1-4 \ 2-6 \ 3-5\end{tabular}} \\ \hline
$[t_0+39+50k, t_0+40+50k]$ & \textbf{\begin{tabular}[c]{@{}c@{}}1-5 \ 2-4 \ 3-6\end{tabular}} \\ \hline
$[t_0+49+50k, t_0+50+50k]$ & \textbf{\begin{tabular}[c]{@{}c@{}}1-6 \ 2-3 \ 4-5\end{tabular}} \\ \hline
\end{tabular}
\caption{\centering Measurement schedule, where $t_0$ is the starting epoch and $k \in \mathbb{Z}^+$ \cite{stuart2017formation} \label{tab:meassched1}}
\end{table}

\subsubsection{Extended Kalman Filter}

The navigation algorithm used to evaluate the performance of this method consists of a simple EKF (Extended Kalman Filter). The purpose of the EKF is to provide an estimate of the system state $x \in \mathbb{R}^p$, following a dynamic system $\Dot{x}(t) = f(x(t),t,w)$ with observations $y(t) = h(x(t),t,\nu)$, in which $y \in \mathbb{R}^m$ describes the available observations,  $w \in \mathbb{R}^p$ represents process noise to account for the dynamic system's modelling innaccuracies and $\nu \in \mathbb{R}^m$ represents measurement noise. Both process and measurement noise are modeled as centered white noise, with $w \sim \mathbb{N}(0,Q)$ and $\nu \sim \mathbb{N}(0,R)$.

\subsubsection*{Prediction model}

In \cite{stuart2017formation}, the assumption is made that no information is available on the spacecraft's initial absolute states in the ECI (Earth Centered Inertial) frame. For that reason, the filter's prediction model follows a PVA (Position-Velocity-Acceleration) model, such that the state vector accounts for the relative position, velocity and acceleration of the spacecraft in the formation.

Let us consider $\delta r_{j/1}$ the position vector of spacecraft $j \in [2,...,6]$ with respect to spacecraft $1$ in an inertial frame centered on spacecraft $1$. Its associated velocity vector is $\delta v_{j/1}$, and its acceleration vector is $\delta a_{j/1}$. The state propagation equation in the PVA model is described as
\begin{equation}
\begin{split}
\Dot{x}_j & = A_jx_j + G_j w_j \Leftrightarrow\\
\begin{bmatrix}
\Dot{\delta r}_{j/1} \\
\Dot{\delta v}_{j/1} \\
\Dot{\delta a}_{j/1}
\end{bmatrix} &= 
\begin{bmatrix}
\mathbf{0} & I & \mathbf{0} \\
\mathbf{0} & \mathbf{0} & I \\
\mathbf{0} & \mathbf{0} & \mathbf{0}
\end{bmatrix}
\begin{bmatrix}
\delta r_{j/1} \\
\delta v_{j/1} \\
\delta a_{j/1}
\end{bmatrix} + \begin{bmatrix}
\mathbf{0} \\
\mathbf{0} \\
I
\end{bmatrix}w_j
\end{split}\label{•}
\end{equation}
where $w_j$ is a centered Gaussian white noise process with covariance $E\{w_j(t)w_j^T(\tau)\} = Q_j\delta(t-\tau) = q_jI\delta(t-\tau)$, and in \cite{stuart2017formation} $q_j$ was set to $(\SI{1e-7}{})^2\SI{}{(\m/\second^3)^2} \quad \forall j \in \{2,\dots,6\}$. The full state vector $x$ for the formation is the concatenation of each relative spacecraft motion state vector $x_j$ for $j \in \{2,\dots,6\}$. The full state dynamic model is therefore
\begin{equation}
\Dot{x} = Ax + G w,
\label{eq:dynmodfull}
\end{equation}
where $A=\mathrm{diag}(A_2,\dots,A_6)$, $G=\mathrm{diag}(G_2,\dots,G_6)$, $x=\{x_j\}_{j=2}^6$ and $w$ is the concatenation of the white noise processes $w_j$, akin to that of the full state vector. The full process covariance matrix $Q$ is similarly the matricial direct sum of all $Q_j, \ \forall j \in \{2,\dots,6\}$. In discrete-time, due to the linear nature of the model, \textit{a priori} state propagation has a closed form solution given by
\begin{equation}
\hat{x}_k = \Phi_{k|k-1}\hat{x}_{k-1},
\label{eq:dynmodfulldisc}
\end{equation}
whereas state covariance matrix propagation follows the equation
\begin{equation}
P_k = \Phi_{k|k-1}P_{k-1}\Phi_{k|k-1}^T + Q_d,
\label{eq:dynmodfulldisc2}
\end{equation}
with the STM (state transition matrix) from states $x$ at time $t_{k-1}$ to time $t_k$ $\Phi_{k|k-1}$ described as 
\begin{equation}
\Phi_{k|k-1} \equiv \Phi_{k}(t_{k-1}) = e^{A(t_k-t_{k-1})}.
\label{eq:stmdiscfilt1}
\end{equation}

The discretized process noise covariance matrix $Q_d$ is calculated according to 
\begin{equation}
Q_d = \int^{t_k}_{t_{k-1}} \Phi_{k}(\tau)GQG^T\Phi_{k}(\tau)^Td\tau,
\label{eq:discqd}
\end{equation}
which has a closed-form solution~\citep{zarchan2013fundamentals}.

\subsubsection*{Observation model}

Relative position measurements will be made between any pair of two spacecraft, here indexed by $j$ and $n$. The range and bearing measurements within a pair are functions of the relative position vector in the inertial frame from one spacecraft to the other:
\begin{equation}
 \delta r_{j/n} = [\delta r_{x,j/n} \ \delta r_{y,j/n} \ \delta r_{z,j/n}]^T.
 \end{equation}
Relative positions between a pair that does not include the chief spacecraft (index $1$) can be described from the relative position vectors that belong to the state vector $x$ as  
\begin{equation}
	\delta r_{j/n}(t) = \delta r_{j/1}(t) - \delta r_{n/1}(t), \forall j \neq n \in \{2,...,6\}.
\end{equation}
Range and bearing (expressed through right ascendancy and declination angles) of spacecraft $j$ from spacecraft $n$ are expressed as 
\begin{align}
	\rho_{j/n}(t) &= \norm{\delta r_{j/n}(t)} + \nu_{\rho}(t) \\
	\psi_{j/n}(t) &= \arctan\left(\frac{\delta r_{y,j/n}(t)}{\delta r_{x,j/n}(t}\right) + \nu_{\psi}(t) \\
	\theta_{j/n}(t) &= \arcsin\left(\dfrac{\delta r_{z,j/n}(t)}{\norm{\delta r_{j/n}(t)}}\right) + \nu_{\theta}(t)
\label{eq:measmodel}
\end{align}
where $\nu_{\rho}(t)$, $\nu_{\psi}(t)$ and $\nu_{\theta}(t)$ are Gaussian white noise processes with covariances set in \cite{stuart2017formation} as $E\{\nu_{\rho}(t)\nu_{\rho}(\tau)\} = (1/3)^2\delta(t-\tau) \SI{}{\m^2}$ and $E\{\nu_{\psi}(t)\nu_{\psi}(\tau)\} = E\{\nu_{\theta}(t)\nu_{\theta}(\tau)\} = (35)^2\delta(t-\tau) \text{arcsec}^2$. These are incorporated into the diagonal entries of the measurement noise covariance matrix $R$. 

The \textit{a posteriori} states and respective covariance matrix (denoted $\hat{x}^{+}$ and $P^{+}$ as opposed to the \textit{a priori} $\hat{x}^{-}$ and $P^{-}$) are computed according to
\begin{align}
	K_k &= P_k^{-}H_k^T(H_kP_k^{-}H_k^T + R)^{-1} \\
	\hat{x}_k^{+} &= \hat{x}_k^{-} + K_k(y_k - \hat{y}_k) \\
	P_k^{+} &= (I - K_kH_k)P_k^{-},
\label{eq:statecovupd}
\end{align}
in which $H_k = \dfrac{dh}{dx}\biggr|_{\hat{x}_k^{-}}$ is the observation matrix (detailed in \citep{stuart2017formation}) and $y_k$ and $\hat{y}_k$ are the real and estimated observations at time $t_k$.

Since the filter is expected to provide state estimates every second, it outputs the \textit{a priori} estimates instead of the \textit{a posteriori} ones for the time steps with no available measurements.

\subsection{Proposed solution}
\label{subsec:propsol}

The previously described solution pertains to a worst-case scenario where no information is available on the formation's initial position in the ECI frame. This is a worst-case scenario assumption, as some information is always expected to be available from launch to the start of mission operations. Assuming that it is, then the absolute state of the chief spacecraft may be incorporated into the state vector, as well as orbital dynamics into the filter's prediction model. The relative position history of the deputy spacecraft should make it possible to correct the chief spacecraft's absolute states and avoid divergence.

The propagation of the absolute position is done considering a simple keplerian model:
\begin{equation}
	\begin{cases}
		\Dot{r}_1 &= v_1 \\
		\Dot{v}_1 &= -\mu\dfrac{r_1}{\norm{r_1}^3} + \omega_1
	\end{cases}.\label{eq:propabs}
\end{equation}
The relative states of spacecraft $j$ with respect to the chief spacecraft $1$ in a local inertial frame centered on the latter, in turn, are propagated according to the following equations:
\begin{equation}
	\begin{cases}
		\Dot{\delta r}_{j/1} &= \delta v_{j/1} \\
		\Dot{\delta v}_{j/1} &= -\mu_{\oplus}\left(\dfrac{r_1 + \delta r_{j/1}}{\norm*{r_1 + \delta r_{j/1}}^3} - \dfrac{r_1}{\norm{r_1}^3}\right) + \omega_j
	\end{cases}.\label{eq:proprel}
\end{equation}
Unlike the PVA model, this propagation model is nonlinear. In \citep{cachim2020}, these equations were computed numerically using \emph{MATLAB}'s \emph{ODE45} solver. However, since this solver may be too computationally heavy to use onboard a NanoSat mission, and since the time steps are relatively short ($\SI{1}{\second}$), these equations may instead be discretised using Euler's method (as described in \citep{zarchan2013fundamentals}):
\begin{equation}
\hat{x}_k = \hat{x}_{k-1} + f(\hat{x}_{k-1})(t_k-t_{k-1}),
\end{equation}
with the STM $\Phi_{k|k-1}$ calculated as
\begin{equation}
\Phi_{k|k-1} = I + F(t_k-t_{k-1}),
\end{equation}
with $F = \dfrac{df}{dx}\biggr|_{\hat{x}_{k-1}}$. The STM is then integrated in Eq. \eqref{eq:discqd} to obtain $Q_d$ and used in Eq. \eqref{eq:dynmodfulldisc} to propagate the state covariance matrix forward each time step. The process noise is modelled similarly to the previous filter, with the exception that the values of $q_1$ and the remaining $q_j$ were manually tuned to $(\SI{1e-6}{\km/\s^2})^2$ and $(\SI{1e-9}{\km/\s^2})^2$, respectively, in order to optimize the filter's performance.

\section{Observability Optimization}
\label{sec:ObsOpt}

Previous studies concluded that the orbital configuration of the spacecraft has a great impact on the system's observability~\citep{psiaki1999autonomous}.
In order to further explore the potential of the proposed method, the orbital configuration for a new spacecraft in the formation  that optimizes the system's navigation performance is now investigated.


The generic formulation for the optimization problem in question is
\begin{align*}
\underset{x \in \mathcal{D}}{\text{minimize}} \quad &f(x) \\
\text{subject to} \quad & f_i(x) \leq 0, \ i\in  \{1,...,m\} \\
& h_j(x) = 0, \ j\in  \{1,...,p\}
\end{align*}
where $x$ describes the initial orbital configuration of the new spacecraft in Classical Orbital Elements, $\mathcal{D}$ is the domain of $x$, $f(x)$ is the objective function that quantifies the observability/performance of the system, and $f_i(x)$ and $h_j$ are the inequality and equality constraints, respectively. The configurations of the objective function  are discussed next in Section \ref{subsec:objfunc}. The optimization variables $x$, their domain $\mathcal{D}$ and respective constraints are discussed in Section \ref{subsec:conssearch}.

\subsection{Objective function}
\label{subsec:objfunc}
Several choices exist regarding the choice of objective function. Two tools were used to construct these objective functions, that serve to evaluate and quantify the observability/performance of an observed dynamic system:

\begin{itemize}
	\item The continuous-time \textit{Observability matrix} $\mathcal{O}$ up to order 3, as described in \ref{app:obsmat}. Within the context of this study, it will be used to study which results would optimize local observability across the trajectory, in the absence of measurement noise;
	\item The inverse of the SFIM (\textit{Standard Fisher Information Matrix}) serves as a lower bound of the state covariance of a discrete-time linear (or linearized) system \cite{rafieisakhaei2018use}. Because the computation of the SFIM for this problem can be numerically innacurate due to the large disparity in state observability, its square root form is used instead, calculated according to Eq. \eqref{eq:srsfim} (following the approach described in \cite{psiaki1999autonomous}), in which $H_k = \dfrac{dh}{dx}\biggr|_{x_k}$ is the observation matrix at time $t_k$ and $\Phi_{k|0}$ the STM (State Transition Matrix) from the states at time $t_0$ to the states at time $t_k$.
\end{itemize}

\begin{equation}
\mathcal{I}=[(R^{-1/2}H_0\Phi_{0|0})^T \ \dots \ (R^{-1/2}H_k\Phi_{k|0})^T]^T
\label{eq:srsfim}
\end{equation}

Two metrics for each of these matrices were considered for optimization: the smallest singular value and the condition number (ratio of largest to smallest singular value):

\begin{itemize}
	\item Maximizing the smallest singular value (or minimizing its negative value, which will be referred to as \textit{Local Unobservability Index} or LUI), which equates to increasing the observability of the least observable state subspace in the context of the observability matrix, or reducing its estimation error variance in the context of the SR-SFIM;  
	\item Minimizing the condition number (or its negative reciprocal value) should lead to a better conditioned matrix, decreasing the disparity in observability or estimation error between the least and most observable subspaces.
\end{itemize}

 These metrics can be taken from any matrix and used to evaluate how close to singular it is. Because the largest singular value of the observability matrix in question in this scenario is always equal to one, both the CN and LUI optimization of this matrix lead to similar results. For this reason, only one of the observability matrix optimized configurations (the LUI) is discussed here. 
 
 While the SR-SFIM is calculated considering a set of discrete observations from a given trajectory, the continuous-time observability matrix is associated with a set of states at a given point in time and space. So, in order to evaluate the observability of a trajectory using the observability matrix in question, its corresponding metric was averaged across time steps.
 
\subsubsection*{Relative Positioning System}

Two different measurement systems for the new spacecraft were considered in this optimization problem: 

\begin{itemize}
 \item The RF/vision-based system described in Section \ref{subsect:rfvis}, with equivalent error model;
 \item An RF-only system, that uses multiple receivers and TOA (Time-Of-Arrival) differencing to estimate the (AOA) Angle-Of-Arrival of the signal, which can be used with absolute orientation knowledge to estimate relative bearing.
\end{itemize}

In both systems, measurement noise is modelled with centered white noise. The values for the standard deviation of the RF/Vision-based system have been described in Section \ref{subsect:rfvis}. For the RF-only system, the ranging measurement noise standard deviation is similar, but the relative bearing is more innacurate, with a standard deviation of $1^{\circ}$ for the right ascension and declination angles (based on the precision presented by the FFRF system onboard the PRISMA mission \cite{harr2006rf}). 

While the PRISMA mission's FFRF system was only operated up to a range of $\SI{30}{\km}$~\cite{harr2006rf}, in this study the assumption is made that the new spacecraft  is designed with suitable \textit{Ad-Hoc} RF-transmitting capabilities such that it can perform measurements regardless of the distance between the spacecraft within the Earth's sphere of influence. 
The RF/vision-based system, in turn, will present a maximum distance constraint to account for the camera's functional range. 
With these assumptions, the results will allow for a more interesting comparison between two  measurement systems that either have greater accuracy or allowed range of motion.

Due to the maximum distance constraint imposed on the RF/vision-based system, the orbital period of any spacecraft employing this system must be similar to that of the chief spacecraft, as otherwise the new spacecraft would eventually drift out of range of the formation. 

Because the objective function only describes the performance of the system for a finite period of time, the relative motion of the spacecraft within the considered time frame needs to be periodical for the performance to be optimized beyond the studied period. For this reason, the orbital period of the new spacecraft with the RF-only system is also kept similar to that of the chief spacecraft.

Because the goal of the observability matrix optimization is only to help visualize the impact of the orbital configuration on the system's observality, they do not present an orbital period constraint. For these results, the optimized period corresponds to that of the greater orbital period between the new and chief spacecraft.



\subsubsection*{Dynamic system approximations}

In order to reduce the computation time of each $f(x)$ evaluation and allow for a more in-depth examination of the search space, some approximations were made regarding the dynamic system. The formation was reduced to a 2 spacecraft system, with one spacecraft being placed at the chief orbit around which the formation was designed and the other being the new spacecraft. Furthermore, the sampling period was changed from the schedule described in Table \ref{tab:meassched1} to a fixed sampling period of $\SI{90}{\second}$, leading to a total of 1000 samples over one orbital period of the chief spacecraft.

The trajectory of states and STM are propagated from the initial variables $x$ with the dynamic equations described in Eqs. \eqref{eq:propabs} and \eqref{eq:proprel} using \emph{MATLAB}'s \emph{ode45}~\citep{shampine1997matlab} solver, with relative and absolute tolerances of $1\mathrm{e}{-3}$ and $1\mathrm{e}{-6}$, respectively. The states are then used to obtain the observation matrices $H_k$, state transition matrices $\Phi_{k|0}$ for the computation of the SR-SFIM according to Eq. \eqref{eq:srsfim} and the continuous-time observability matrices (according to Eq. \eqref{eq:obsmatrelstat}).

\subsubsection*{Eclipse condition}

Within a configuration in which the new spacecraft has a wider search space available, it is possible for the two spacecraft to stand on opposite sides of the Earth, blocking the measurement system's field-of-view. This obstruction is accounted for in the objective function, such that measurements become unavailable during the \enquote{eclipse} period. Whenever the new spacecraft stands in a cylindrical shadow zone behind the planet opposite to the chief spacecraft, the local observability matrix LUI and the negative reciprocal of the CN are set to zero. In the SR-SFIM scenarios, the observation matrices $H_k$ corresponding to the eclipsed time samples are set to zero.

To further discourage the presence of occultation periods in these results (to avoid the positioning error drift from propagation-only estimation), the SR-SFIM-related cost functions are multiplied by the ratio of uneclipsed to total observations used in the functions calculation.

\subsection{Constraints and Search Domain}
\label{subsec:conssearch}
The optimization variables, $x$, describes the initial state of the new spacecraft. Classical Orbital Elements are generically used to represent it ($x=\{a_2,e_2,i_2,\Omega_2,\omega_2,\nu_2\}$, where $a$ is the SMA, $e$ is the eccentricity, $i$ the inclination, $\Omega$ the RAAN, $\omega$ the argument of the perigee and $\nu$ the true anomaly). The subscript $2$ refers to the new spacecraft, whereas the subscript $1$ refers to the chief spacecraft.

The domain $\mathcal{D}$ of $x$ needs to be constrained, however. 
The perigee of the orbit must stay above a given threshold to avoid excessive atmospheric drag (defined at $\SI{6678}{\km}$).
For the RF-only configurations, since the new spacecraft's orbital period is equal to that of the chief (and, therefore, the semi-major axis $a_2=a_1=a$), these restrictions limit the eccentricity such that $e_2 \in [0,0.85]$. 

For the observability matrix optimization study, which presents no orbital period constraint, nonlinear constraints are avoided by replacing the elements $a_2$ and $e_2$ with the radius of the orbit at the apsides $r_1,r_2 \in [R_{min}, R_{max}]$. $a_2$ and $e_2$ are then calculated from these elements according to $a_2 = (r_1 + r_2)/2$ and $e_2 = |r_1-r_2|/(r_1+r_2)$. Both $r_1$ and $r_2$ must remain between the lower limit of $\SI{6678}{\km}$ and the upper limit of $\SI{3e5}{\km}$, defined to limit the computation time of the objective function (proportional to the greater orbital period of the two spacecraft).

The functional range restriction for the vision-based system, however, requires a nonlinear constraint. Setting this distance as $d_{max}=\SI{480}{\km}$ (based on the PRISMA mission's VBS system with a $\SI{20}{\km}$ margin), it is not easy to apply this restriction on the initial states of the formation. In order to do so, relative motion was modelled by the approximated equations in the LVLH (Local-Vertical Local-Horizontal) frame derived from the Hill-Clohessy-Wiltshire equations. Given that the chief orbit is considered equatorial and circular, and assuming that both spacecraft have the same orbital period ($a_2=a_1=a$), the relative motion of the new spacecraft around the chief spacecraft can be approximately described in a LVLH frame from the keplerian elements as follows \citep{ou2018observability}:

\begin{equation}
	\begin{cases}
		\delta x_{2/1} \simeq -ae_{2}\cos(\omega_1 + M_1-\omega_{2}) \\
		\delta y_{2/1} \simeq a[\omega_2 + M_2 - \omega_1 - M_1 + \Omega_{2}-\Omega_1  \\  + 2e_{2}\sin(\omega_1 + M_1-\omega_{2})] \\
		\delta z_{2/1} \simeq ai_{2}\sin(\omega_1 + M_1)
	\end{cases}\label{eq:relmotionsimp}
\end{equation}

Based on this simplified model, the following approximated formula for the maximum distance constraint between the spacecraft was designed:
\begin{multline} 
 		a\bigl[(|\omega_2 + M_2 + \Omega_2 - \omega_1 - M_1 - \Omega_1| + 2e_2)^2\\ + (i_2-i_1)^2\bigr]^{1/2} - d_{max} < 0 \label{eq:visionconstraint}
\end{multline}
with $M$ the mean anomaly, which can be obtained from the true anomaly $\nu$ and eccentricity $e$. This formulation is built on the generally conservative assumption that the maximum distance is always met with a simultaneous maximum separation in the along-track and cross-track directions ($\delta y_{2/1}$ and $\delta z_{2/1}$ in Eq. \eqref{eq:relmotionsimp}, respectively).

\subsection{Optimization Results}


For the optimization problems with nonlinear constraints, \emph{MATLAB}'s \emph{fmincon} Interior Point Algorithm with multi-start was used as a global solver \cite{byrd2000trust}. 
For the remaining ones, \emph{PSwarm} was used \cite{vaz2007pswarm}. Each algorithm is run for a period of $\sim 8$ hours. The reader is referred to \citep{cachim2020} for more information on the configuration of the optimization problems and respective solvers. The optimized results are shown in Table \ref{tab:optobsmatorbs} and analysed next.

\begin{table*}[tb]
\centering
\resizebox{\textwidth}{!}{
\begin{tabular}{lllccccccc}
\hhline{==========}
\multicolumn{3}{l}{\multirow{2}{*}{\textbf{Objective  function}}} &
  \multicolumn{6}{c}{\textbf{New SC orbit states} $\bm{x_{opt}}$} &
  \multirow{2}{*}{\textbf{f(x)}} \\
\multicolumn{3}{l}{} &
  $\bm{a (km)}$ &
  $\bm{e}$ &
  $\bm{i (deg)}$ &
  $\bm{\Omega (deg)}$ &
  $\bm{\omega (deg)}$ &
  $\bm{\nu (deg)}$ &
   \\ \hline
\multicolumn{2}{l}{\textbf{Observability Matrix}} & \textbf{LUI} & $6678$  & $0$       & $88.34$    & $129.1$ & $351.5$ & $121.2$ & $-9.905\mathrm{e}{-7}$  \\
\multirow[t]{4}{*}{\textbf{SR-SFIM}} &
  \multirow[t]{2}{*}{\textbf{RF/Vision}} &
  \textbf{CN} &
  $43399$ &
  $5.203\mathrm{e}{-4}$ &
  $0.629$ &
  $195.3$ &
  $354.9$ &
  $169.9$ &
  $-2.578\mathrm{e}{-9}$ \\
 &                                   & \textbf{LUI} & $43399$ & $1.714\mathrm{e}{-3}$ & $0.404$ & $208.2$   & $175.4$ & $336.5$ & $-4.276$     \\
 & \multirow[t]{2}{*}{ \textbf{RF-only}} & \textbf{CN}  & $43399$ & $8.439\mathrm{e}{-1}$ & $89.46$    & $198.2$   & $284.8$ & $190.1$ & $-9.310\mathrm{e}{-12}$ \\
 &                                   & \textbf{LUI} & $43399$ & $3.632\mathrm{e}{-1}$ & $18.62$ & $90.79$   & $149.9$ & $119.4$ & $-4.173\mathrm{e}{-2}$  \\ \hhline{==========}
\end{tabular}
}
\caption{\centering Initial COE of the optimized new spacecraft configurations.}
  \label{tab:optobsmatorbs}
\end{table*}

\subsubsection{Observability Matrix}

\begin{figure}[tb]
  \centering
  \includegraphics[width=.9\linewidth]{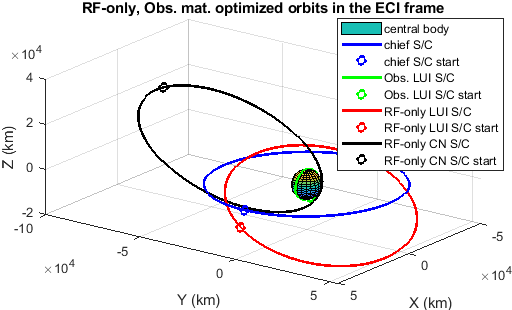}
  \caption{\centering Observability matrix and RF-Only SR-SFIM optimized configurations. The dots mark the initial S/C states.}
  \label{fig:opt_rf_fixed_obs_cnlui}
\end{figure}

The observability-optimized orbital configuration for the new spacecraft is shown in Figure \ref{fig:opt_rf_fixed_obs_cnlui}. The obtained new SC orbit is near polar ($i_{new} \sim 90^{\circ}$), and at the lowest allowed altitude. This result is in accordance with the conclusions drawn in \citep{ou2018observability} on how the observability of this system improves with greater magnitudes of in-track and cross-track distances.

\begin{figure}[tb]
  \centering
  \includegraphics[width=.9\linewidth]{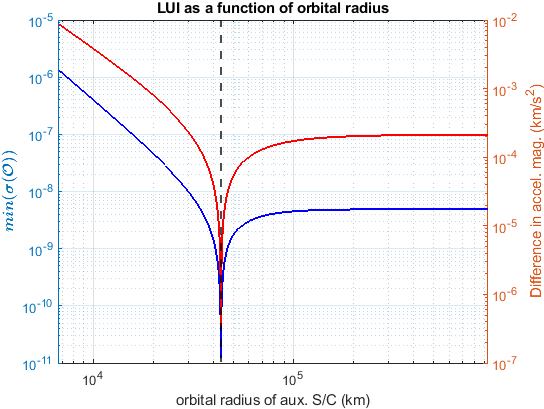}
  \caption{\centering Observability matrix and RF-Only SR-SFIM optimized configurations. The dots mark the initial S/C states.}
  \label{fig:LUI_orbrad}
\end{figure}

In order to provide some reasoning behind the obtained results, we seek to visualize how the difference in orbital radius affects observability.
Noting that the observability matrix evaluates observability locally (at a specific point in time rather than a state trajectory), we take the initial states of the chief virtual orbit and consider that at a given point in time the deputy spacecraft’s absolute position vector is aligned with the chief orbit’s absolute position vector.  The LUI of the observability matrix is evaluated for different values of the deputy spacecraft’s orbital radius (we consider the deputy’s velocity to be adjusted such that its orbit is always circular and equatorial). Taking 10000 samples ranging from the edge of the Earth’s atmosphere up to the edge of the Earth’s gravitational sphere of influence ($\sim \SI{9.26e+05}{\km}$ \citep{vallado2001fundamentals}), the plot in Figure \ref{fig:LUI_orbrad} was drawn.

Based on the observations made in \citep{psiaki1999autonomous} on how the observability of the system is connected to the gravity gradient tensor, the difference in magnitude of gravitational acceleration was added to the plot in Figure \ref{fig:LUI_orbrad} in order to observe whether its curve would be similar to that of the LUI. The vertical dashed line marks the orbital radius of the chief orbit. 
These results help to understand why the optimized orbit would maximize the observability of the system. In general, these results seem to indicate that large differences in magnitude of gravitational acceleration are the key factor affecting the local observability of the least observable (absolute) states.
This hypothesis can be derived from the formula of the critical entries of the observability matrix defined by the difference in gravity gradient tensors $G_1 - G_0$ described in Eqs. \eqref{eq:obsmatrelstat} and \eqref{eq:Gi}.
 These results support this hypothesis because the optimized auxiliary orbits are those which maximize this difference in magnitude, as evidenced by its plot in Figure \ref{fig:LUI_orbrad}. They also indicate that increasing the maximum orbital radius of the search space of the problem from $\SI{3e+05}{\km}$ to the edge of the Earth gravity sphere of influence would likely not have resulted in a lower cost function value, since Figure \ref{fig:LUI_orbrad} shows the smallest singular value being largest at the lower bound of the orbital radius of the new spacecraft.

\subsubsection{SR-SFIM RF/vision system}

\begin{figure}[tb]
  \centering
  \includegraphics[width=.7\linewidth]{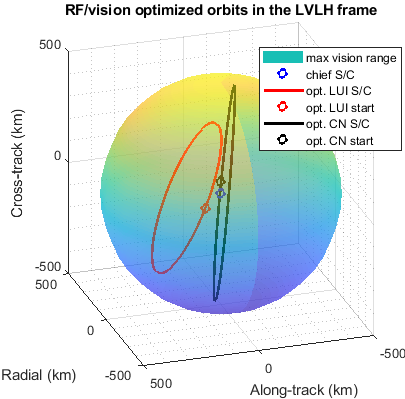}
  \caption{\centering Optimized orbits for the RF/vision based system in the LVLH frame centered around the chief orbit.}
  \label{fig:optvisorbslvlh}
\end{figure}

We now analyse the CN and LUI SR-SFIM optimization results for the RF ranging and vision bearing measurement system for the new spacecraft. The optimized configurations are shown in Figure \ref{fig:optvisorbslvlh} in the LVLH plane, since their proximity to the chief spacecraft makes them indiscernible in the ECI frame.

Unlike in the case of the observability matrix, noticeable differences can be observed between the optimization of the CN and the LUI of the SR-SFIM. Since the state error covariance of the relative states increases with the distance between the spacecraft, all singular values are affected by the orbital configuration, unlike with the observability matrix. 

The CN-optimized configuration presents a new relative orbit with wider out-of-plane motion and closer in-plane motion when compared to the LUI-optimized configuration.

By performing singular value decomposition of the SR-SFIM matrix of the optimized results, we can use the right singular vectors to know which states the smallest and largest singular values are most associated with (reminding the reader that the SR-SFIM is associated with the states in an inertial cartesian coordinate system as with the EKF, rather than COE).
 In both configurations, the largest singular value is linked with $\delta v_y$, the Y component of the relative velocity of the new spacecraft with respect to the chief orbit in the ECI frame, while the smallest singular value corresponds most to $r_y$, the Y component of the chief spacecraft’s absolute position vector in the ECI frame. 

These results coincide with those of the observability matrix in the sense that more accurate information is available on the relative states than the absolute ones.
The singular values of the SR-SFIM corresponding to the relative states are expected to decrease with distance due to the fixed angular error of the relative observations. On the other hand, as is shown in Figure \ref{fig:LUI_orbrad}, greater differences in orbital radius will increase the observability of the absolute states.

\begin{figure}[tb]
  \centering
  \includegraphics[width=.8\linewidth]{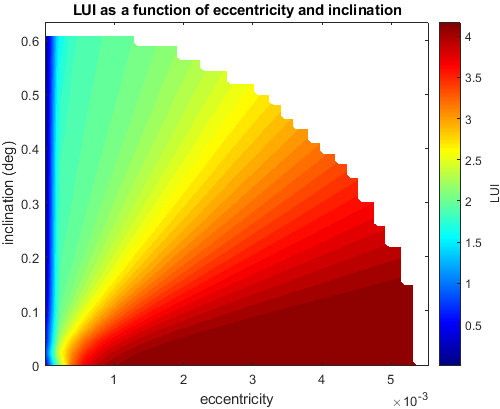}
  \caption{\centering LUI of $\mathcal{I}$ with the RF/vision-based system as a function of inclination and eccentricity.}
  \label{fig:optrforbseffectlui}
\end{figure}

\begin{figure}[tb]
  \centering
  \includegraphics[width=.8\linewidth]{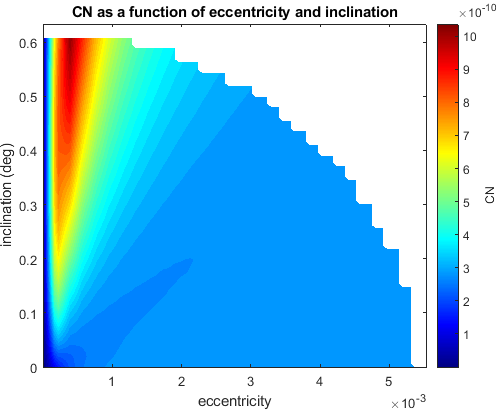}
  \caption{\centering CN of $\mathcal{I}$ with the RF/vision-based system as a function of inclination and eccentricity.}
  \label{fig:optrforbseffectcn}
\end{figure}

In order to observe how these different factors affect the LUI and CN of the SR-SFIM, we calculate the values of these objective functions for a new spacecraft placed on orbits with varying degrees of inclination and eccentricity. The remaining orbital elements of the new spacecraft are kept similar to those of the chief orbit. The results in Figures  \ref{fig:optrforbseffectlui} and \ref{fig:optrforbseffectcn} were produced by sampling the objective function $30\times30$ grid for both the CN and the LUI, and linearly interpolating these samples for a \enquote{smoother} plot.

Figure \ref{fig:optrforbseffectlui} shows that the absolute value of the LUI generally decreases with the inclination and increases with the eccentricity up until $e \sim \SI{1.5e-03}{}$, where it appears to plateau. Higher eccentricity of the new spacecraft's orbit cause greater differences in orbital radius, which may explain the positive impact of the eccentricity on the LUI. 

Because the information available on the absolute states comes from the estimation of the relative states, we should expect the accuracy of the information on the absolute states to be linked to that of the relative states. If greater interspacecraft distances lead to more inaccurate relative state estimation, so should absolute state estimation be negatively impacted by greater distances. This may justify why greater degrees of inclination of the new spacecraft's orbit would lower the absolute value of the LUI, and why the latter plateaus at higher eccentricities in Figure \ref{fig:optrforbseffectlui}. Although this contradicts the level of inclination and eccentricity presented by the LUI-optimized orbit, it is important to note that the effect of the remaining orbital parameters is not considered in this plot.

Figure \ref{fig:optrforbseffectcn} shows that the CN is lowest at a low eccentricity and high inclination orbit. As previously mentioned, minimizing the CN (or maximizing its absolute value) is equivalent to shortening the difference in accuracy of information available between the least and most observable state subspaces in order to achieve a better conditioned estimation problem. In this context, it is equivalent to maximizing the information available on the least observable absolute position states while minimizing that of the most observable state ($\delta v_y$ in both optimized configurations). It is therefore possible to infer the look of the equivalent plot for the most observable state subspace based on the plots in Figures \ref{fig:optrforbseffectlui}   and \ref{fig:optrforbseffectcn}.

\subsubsection{SR-SFIM RF-only system}

Finally, in this section, we seek to evaluate the optimized orbital configurations for the SR-SFIM RF-only system. The results are shown in Figure \ref{fig:opt_rf_fixed_obs_cnlui}.

None of the optimized new orbits present Earth-eclipsed periods. Whereas the CN-optimized configuration is highly eccentric and near-polar, the LUI-optimized orbit presents a comparatively small degree of eccentricity ($e_2 \simeq 0.37$) and inclination ($i_2 \simeq 19^{\circ}$). 

Once again, we perform singular value decomposition on the optimized SR-SFIMs. The state most associated with the smallest singular value for the LUI-optimized configuration is once again $r_y$, whereas for the CN-optimized one it is $\delta r_x$. The state most linked with the largest singular value, in turn, is $v_y$ for the CN configuration and $\delta v_y$ for the LUI configuration.

The fact that the most observable state for the CN-optimized configuration is an absolute state of the chief orbit appears to indicate that, for orbits far enough apart, the accuracy of the information available on the relative states can become worse than the one available on the absolute states.

\begin{figure}[tb]
  \centering
  \includegraphics[width=.8\linewidth]{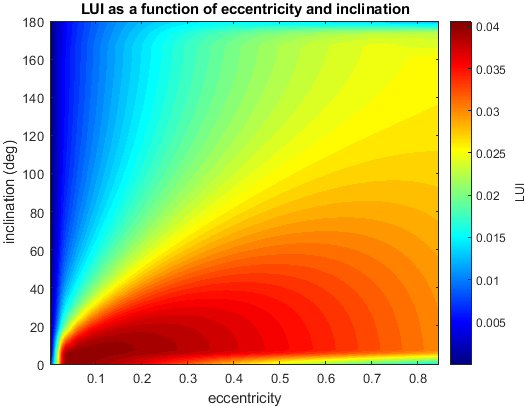}
  \caption{\centering LUI of $\mathcal{I}$ with the RF-only system as a function of inclination and eccentricity.}
  \label{fig:optrforbseffectluirf}
\end{figure}

\begin{figure}[tb]
  \centering
  \includegraphics[width=.8\linewidth]{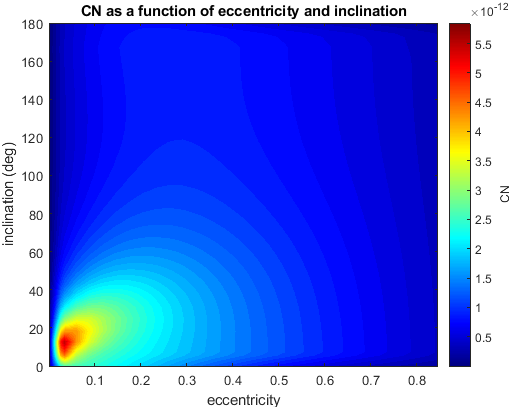}
  \caption{\centering CN of $\mathcal{I}$ with the RF-only system as a function of inclination and eccentricity.}
  \label{fig:optrforbseffectcnrf}
\end{figure}

Analogous plots to those of Figures \ref{fig:optrforbseffectlui} and \ref{fig:optrforbseffectcn} are shown in Figures \ref{fig:optrforbseffectluirf} and \ref{fig:optrforbseffectcnrf} in the context of the RF-Only system optimization problems. 

For low values of eccentricity and inclination ($< 0.1$ and $<14^{\circ}$, respectively) in Figures \ref{fig:optrforbseffectluirf} and \ref{fig:optrforbseffectcnrf}, the plots of both CN and LUI appear to show a similar shape to the plots of their RF/Vision-based system counterparts in Figures \ref{fig:optrforbseffectlui} and \ref{fig:optrforbseffectcn}. 
It is important to note once again that these plots do not present the obtained optimal values, since the remaining orbital elements of the new spacecraft orbit are set equal to those of the chief orbit and not of the optimized solutions.

\section{System Sensor Reduction}
\label{sec:systsensred}

One main disadvantage presented by the RF/Vision-based navigation solution proposed in \citep{stuart2017formation} is the need for constant slewing so that the spacecraft may point their cameras towards each other. Eliminating any redundant relative bearing measurements could help overcome this issue. 

As mentioned in Section \ref{subsubsec:syssensred}, it has been shown in the literature that, in a 2 spacecraft system with keplerian dynamics performing ranging measurements only, 9 out of 12 states may be observable at best. Due to the spherical symmetry of the gravity field, the absolute orientation elements $\omega$,$\Omega$ and $i$ cannot be observed \cite{hill2007autonomous}.

Within the potentially observable subspace are the relative orientation elements $\theta$, $\phi_1$ and $\phi_2$. Of these, $\theta$ is the angle between the orbital planes, $\phi_1$ is the angular distance along the orbit of spacecraft $1$ from the periapsis to one of the two intersections of the orbits, and viceversa for $\phi_2$ with respect to the orbit of the second spacecraft. These parameters are functions of the absolute orientation elements of both spacecraft. Knowing one of the spacecraft's absolute orientation elements, it may be possible to deduce those of the other spacecraft.

The hypothesis we therefore seek to validate is whether, in the context of the SunRISE mission, it is sufficient to have only one pair of spacecraft performing relative positioning measurements while the remaining pairs perform ranging-only for the system to remain fully observable. Since the absolute states of the spacecraft performing full relative positioning measurements should be known, those of the spacecraft performing ranging-only with either of these two spacecraft may also become observable.

\subsection{Observability Analysis}

We wish to validate the observability of the previously described system. In order to do so, we evaluate a simplified system composed of 3 spacecraft, with one chief, and two deputy spacecraft, one of which is the LUI-optimized RF/Vision configuration of the new spacecraft. The states $x$ of the system describe the position and velocity vectors of these spacecraft in the ECI frame. The chief and the deputy spacecraft of the original formation perform ranging-only measurements while the chief and the new spacecraft perform both relative range and bearing measurements simultaneously. These measurements are taken every $\SI{90}{\second}$ over one orbital period of $\SI{25}{\hour}$. The propagation model is similar to the one described in Eqs. \eqref{eq:propabs} and \eqref{eq:proprel}, while the measurement model is described in Eqs. \eqref{eq:measmodel}. We describe this discrete-time observed dynamic system with
\begin{equation}
	\begin{cases}
		x(t_k) \triangleq x_k = \phi(x_{k-1}) &\approx \Phi_{k|k-1}x_{k-1}\\
		y(t_k) \triangleq y_k = h(x_{k}) &\approx H_k x_k
	\end{cases}.\label{eq:discsys}
\end{equation}
in which $\Phi_{k|k-1}$ is the STM from states $x_{k-1}$ to states $x_k$ and $H_k$ the observation matrix at time $t_k$. The state propagation and STM are both calculated using \emph{MATLAB}'s \emph{ODE45} solver with relative and absolute tolerances of $\SI{1e-3}{}$ and $\SI{1e-6}{}$, respectively. With these matrices, we construct the discrete-time observability matrix as
\begin{equation}
\mathcal{O} = \begin{bmatrix}
         H_0\\
         H_1\Phi_{1|0} \\
         \vdots\\
         H_{k}\Phi_{k|0}
    \end{bmatrix}.
\end{equation}

 The resulting matrix has full rank 18, albeit with a condition number of $\SI{5.55e9}{}$. This implies that the system, despite observable in theory, presents a large disparity in state subspace observability and is therefore ill-conditioned. Simulations will be used to better evaluate the performance of this type of system.

\section{Simulation Results}
\label{sec:simres}

In this Section, the navigation performance of the spacecraft formation is evaluated in a simulation environment. This analysis aims to meet the following objectives: 
\begin{enumerate*}
	\item compare the accuracy of the filtering solutions described in Sections \ref{subsect:rfvis} and \ref{subsec:propsol};
	\item evaluate the impact of the different optimized auxiliary spacecraft configurations on the positioning performance of the formation;
	\item help to validate the feasibility of the sensor-reduced system described in Section \ref{sec:systsensred};
\end{enumerate*}
These goals are tackled separately in Sections \ref{subsec:filtcomp}, \ref{subsec:simulres} and \ref{subsec:redsyssimul}, respectively.

\subsection{Orbit determination methods}
\label{subsec:filtcomp}

A Monte Carlo analysis is performed, with $M=40$ trials, in order to compare the performance of the orbit determination methods described in Section \ref{sec:locmet}.

The real trajectory simulation approach in \cite{stuart2017formation} consisted of a point-mass gravitational model with a constant acceleration in an arbitrary direction representing non-keplerian perturbations. In this study, since the spacecraft's states in the ECI frame are being estimated and the method's observability is dependent on their orbital configuration, a more complete model was used. The open-source \emph{ODTBX} Toolbox\footnote{\url{http://odtbx.sourceforge.net/} (Last accessed 05/12/2020)} was used to simulate the true trajectories of the spacecraft, with dynamic models considering solar radiation pressure, gravitational pull from the Sun and Moon and asymmetric Earth gravity geopotential model. Each simulation period covers two orbits ($\sim \SI{50}{\hour}$).

The true trajectories are propagated from the initial set of states of each spacecraft, which are described in \ref{app:EKF}.

 In \cite{stuart2017formation}, an error of $\SI{100}{\m}$ and $\SI{1}{\cm/\s}$ in a random direction was given to the initial relative position and velocity vector state estimates $\hat{\delta r}_{j/1}(t_0)$ and $\hat{\delta v}_{j/1}(t_0)$, respectively, while the initial accelerations were assumed to be zero. The corresponding diagonal entries of the initial state covariance matrix are the squared value of that same initial error ($\SI{0.01}{\km^2}$ for position entries and $\SI{1e-10}{(\km/\s)^2}$ for velocity entries), with the exception of the acceleration entries, which are set to $\SI{1e-14}{(\km/\s^2)^2}$. Save for the acceleration entries, these initialization parameters also apply to the simulation of the new filter.

The RMS (Root Mean Square) error values of the original and proposed filter are compared in Table \ref{tab:filtres}. The PVA EKF results are replicated in the new simulation environment and compared with those obtained in \cite{stuart2017formation}. The RMS values only account for the period after which the filters have converged ($\sim \SI{200}{\minute}$). The mean relative position error is the average of the RMS position error of the deputy spacecraft with respect to the chief spacecraft, whereas the absolute position error corresponds to the RMS error of the absolute position of the chief spacecraft.

\begin{table}[tb]
\centering
\begin{tabular}{lccc}
\hhline{====}
\textbf{Filter} &
  \multicolumn{1}{l}{\textbf{\begin{tabular}[c]{@{}l@{}}Initial abs. \\ pos. err (km)\end{tabular}}} &
  \textbf{\begin{tabular}[c]{@{}c@{}}Abs. pos.\\ error (km)\end{tabular}} &
  \textbf{\begin{tabular}[c]{@{}c@{}}Mean rel.\\ pos. error (m)\end{tabular}} \\ \hline
\textbf{\begin{tabular}[c]{@{}l@{}}PVA EKF\\ (results in \cite{stuart2017formation})\end{tabular}}                  & -     & -     & 3.6  \\
\textbf{\begin{tabular}[c]{@{}l@{}}PVA EKF \\ new simulation\end{tabular}}                & -     & -     & 3.560  \\
\multirow[t]{4}{*}{\textbf{\begin{tabular}[c]{@{}l@{}} \\ Added abs.\\ states EKF\end{tabular}}} & 0.1   & 4.506  & 0.1372 \\
                                                                                  & 10    & 4.589  & 0.1369 \\
                                                                                          & 1000  & 4.721  & 0.1378 \\
                                                                                          & 10000 & 49.222 & 0.4833 \\ \hhline{====}
\end{tabular}
\caption{RF/vision filter RMS error comparison.}\label{tab:filtres}
\end{table}

Both simulation environments lead to similar results for the PVA EKF. The proposed alteration to the filter improves the relative positioning accuracy by a factor of $\sim 25$. 

\begin{figure}[tb]
  \centering
  \includegraphics[width=.8\linewidth]{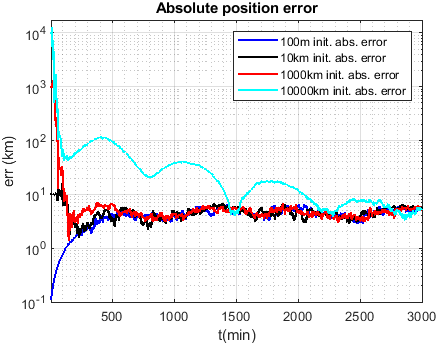}
  \caption{\centering Evolution of the absolute position errors with varying initial chief spacecraft absolute position error.}
  \label{fig:abs_err_noopt_all}
\end{figure}

The proposed filter, however, requires an initial estimate of the absolute position, whereas the PVA filter was designed on the assumption that no such information was available. The proposed solution's robustness to poor initial absolute position error knowledge was also tested. Values of $\SI{10}{\km}$, $\SI{1000}{\km}$ and $\SI{10000}{\km}$ were chosen for the initial chief's absolute position error, with an adjusted initial state covariance. The RMS error values for these MC simulations are displayed in Table \ref{tab:filtres}, and the time evolution of the chief's absolute position error from sample runs are shown in Figure \ref{fig:abs_err_noopt_all}. The results demonstrate the method's robustness to poor initial absolute position information. All of the studied scenarios converged within an orbital period with the exception of the scenario with an initial error of $\SI{10000}{\km}$, in which the filter converges slower.

The proposed modifications to the RF/vision-based navigation solution seem to improve the overall performance of the system, not only increasing the relative positioning accuracy, but also by allowing the formation to autonomously determine its absolute position. Still, its absolute positioning accuracy is considerably worse than that of the chosen GNSS-based navigation solution for the mission (by a factor of $\sim 4000$). Also, compared to the results obtained with the same filter using the \emph{ODE45} solver presented in \citep{cachim2020}, the use of Euler's method degrades the absolute positioning accuracy by a factor of $\sim 2-3$, whereas the relative positioning accuracy is only slightly affected. 

\subsection{Optimized configurations}
\label{subsec:simulres}

We now seek to evaluate the impact of the optimized auxiliary spacecraft on the positioning performance of the formation in a simulation environment. However, in order to do so, a new measurement schedule including the additional spacecraft needs to be designed.

Two sets of results were obtained with different measurement schedules: one in which the same assumptions of the original schedule are made in which no spacecraft can perform measurements with more than one other spacecraft simultaneously, leading to the adapted schedule shown in \ref{tab:meassched2} and denoted as schedule 1; and the other in which the measurements between the new and the chief spacecraft run parallel to the original schedule, leading to the parallel schedule shown in Table \ref{tab:meassched1} and denoted as schedule 2.

\begin{table}[htp]
\centering
\begin{tabular}{|l|c|}
\hline
\multicolumn{1}{|c|}{\textbf{Time Interval, min}} & \textbf{S/C Pairs} \\ \hline
\multicolumn{1}{|c|}{$[t_0+9+70k, t_0+10+70k]$} & \textbf{\begin{tabular}[c]{@{}c@{}}1-2 \ 3-4 \ 5-6\end{tabular}} \\ \hline
$[t_0+19+70k, t_0+20+70k]$ & \textbf{\begin{tabular}[c]{@{}c@{}}1-3 \ 2-4 \ 5-7\end{tabular}} \\ \hline
$[t_0+29+70k, t_0+30+70k]$ & \textbf{\begin{tabular}[c]{@{}c@{}}1-4 \ 2-7 \ 3-6\end{tabular}} \\ \hline
$[t_0+39+70k, t_0+40+70k]$ & \textbf{\begin{tabular}[c]{@{}c@{}}1-5 \ 2-6 \ 3-7\end{tabular}} \\ \hline
$[t_0+49+70k, t_0+50+70k]$ & \textbf{\begin{tabular}[c]{@{}c@{}}1-6 \ 2-5 \ 4-7\end{tabular}} \\ \hline
$[t_0+59+70k, t_0+60+70k]$ & \textbf{\begin{tabular}[c]{@{}c@{}}1-7 \ 3-5 \ 4-6\end{tabular}} \\ \hline
$[t_0+69+70k, t_0+70+70k]$ & \textbf{\begin{tabular}[c]{@{}c@{}}2-3 \ 4-5 \ 6-7\end{tabular}} \\ \hline
\end{tabular}
\caption{\centering Adapted measurement schedule (schedule 1), where $t_0$ is the starting epoch and $k \in \mathbb{Z}^+$ \label{tab:meassched2}}
\end{table} 

The simulations are once again run with the Monte-Carlo method, with $M=40$ samples, with the initial position state estimates placed $\SI{100}{\m}$ away from the real initial position in a random direction. The simulation period is lowered to one orbital period of the original formation ($\sim 25$ hours), since it is sufficient for the filter to achieve convergence. The diagonal entries of the process noise covariance matrix $Q$ corresponding to the new spacecraft are equal to $q_7 = \alpha_{new}(\SI{1e-9}{\km/\second^{2.5}})^2$, with the values of $\alpha_{new}$ for each configuration defined in Table \ref{tab:ekfpnparam} in \ref{app:EKF}. The remaining aspects of the simulations were kept similar to those of Section \ref{subsec:filtcomp}.

Tables \ref{tab:errsched1} and \ref{tab:errsched2} show the RMS error values obtained for each of the optimized configurations with schedules 1 and 2, respectively. The first column describes the absolute positioning error of the chief spacecraft, the second column the mean relative positioning error of the original deputy spacecraft (not counting the optimized auxiliary spacecraft), and finally, the last column describes the relative positioning error of the new spacecraft. The RMS error values obtained for the original formation with no auxiliary spacecraft are also shown. These are evaluated for the first orbit after convergence, as opposed to the $\SI{50}{\hour}$ simulation period of the results shown in Table \ref{tab:filtres}, to ensure the compared results are obtained in similar conditions.

\begin{table}[tb]
\centering
\begin{tabular}{llccc}
\hhline{=====}
\multicolumn{2}{l}{\textbf{\begin{tabular}[c]{@{}l@{}}New SC\\ configuration\end{tabular}}} &
  \textbf{\begin{tabular}[c]{@{}c@{}}Abs. pos. \\ error (km)\end{tabular}} &
  \textbf{\begin{tabular}[c]{@{}c@{}}Mean form. rel.\\ pos. error (m)\end{tabular}} &
  \textbf{\begin{tabular}[c]{@{}c@{}}new SC rel. \\ pos. error (m)\end{tabular}} \\ \hline
\multicolumn{2}{l}{\textbf{None}} & 4.251 & 0.1383 & - \\
\multirow[t]{2}{*}{\textbf{RF/Vision}} & \textbf{LUI} & 4.321 & 0.1655 & 5.754 \\
                                    & \textbf{CN}  & 4.354 & 0.1642 & 4.950 \\
\multirow[t]{2}{*}{\textbf{RF-Only}}   & \textbf{LUI} & 4.239 & 0.1665 & 7315  \\
                                    & \textbf{CN}  & 10.29 & 0.2116 & 2.996$\mathrm{e}{5}$ \\ \hhline{=====}
\end{tabular}
\caption{\centering Mean absolute and relative error for the optimized orbital configurations with schedule 1.\label{tab:errsched1}}
\end{table}

\begin{table}[tb]
\centering
\begin{tabular}{llccc}
\hhline{=====}
\multicolumn{2}{l}{\textbf{\begin{tabular}[c]{@{}l@{}}New SC\\ configuration\end{tabular}}} &
  \textbf{\begin{tabular}[c]{@{}c@{}}Abs. pos. \\ error (km)\end{tabular}} &
  \textbf{\begin{tabular}[c]{@{}c@{}}Mean form. rel.\\ pos. error (m)\end{tabular}} &
  \textbf{\begin{tabular}[c]{@{}c@{}}new SC rel. \\ pos. error (m)\end{tabular}} \\ \hline
\multicolumn{2}{l}{\textbf{None}} & 4.251 & 0.1383 & - \\
\multirow[t]{2}{*}{\textbf{RF/Vision}} & \textbf{LUI} & 4.230 & 0.1381 & 5.568 \\
                                    & \textbf{CN}  & 4.228 & 0.1378 & 5.094 \\
\multirow[t]{2}{*}{\textbf{RF-Only}}   & \textbf{LUI} & 4.068 & 0.1385 & 7478  \\
                                    & \textbf{CN}  & 9.900 & 0.1706 & 2.849$\mathrm{e}{5}$ \\ \hhline{=====}
\end{tabular}
\caption{\centering Mean absolute and relative error for the optimized orbital configurations with schedule 2.\label{tab:errsched2}}
\end{table}

\begin{figure}[tb]
  \centering
  \includegraphics[width=.5\linewidth]{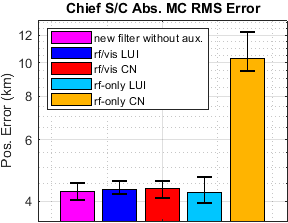}
  \caption{\centering Monte-Carlo averaged  absolute position RMS error results for the configurations with schedule 1, with upper and lower error bounds.}
  \label{fig:barplotsched1}
\end{figure}

\begin{figure}[tb]
  \centering
  \includegraphics[width=.5\linewidth]{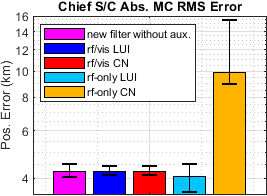}
  \caption{\centering Monte-Carlo averaged  absolute position RMS error results for the configurations with schedule 2, with upper and lower error bounds.}
  \label{fig:barplotsched2}
\end{figure}

To help compare the absolute positioning RMS error results, Figures \ref{fig:barplotsched1} and \ref{fig:barplotsched2} display these in bar plot format for schedules 1 and 2, respectively. The Monte-Carlo averaged absolute error values are shown with the respective error bars indicating the smallest and largest error values obtained in the Monte-Carlo runs.

The results show that the RF/Vision and the LUI-optimized RF-only configurations for the auxiliary spacecraft match the absolute positioning accuracy of the formation, while the CN-optimized RF-only configuration worsens it, regardless of the schedule. While the former three configurations do provide some slightly better performance in absolute positioning with schedule 2, these gains are not substantial when considering the dispersion of the Monte Carlo results.

When comparing the results from both schedules, it is possible to note that schedule 2 generally provides better relative positioning accuracy within the original formation.
The RF/Vision and the LUI-optimized RF-only configurations all present very close absolute and mean formation relative positioning errors for both schedules.

The large errors observed in the RF-only CN-optimized configuration appear to be linked to factors that are unaccounted for in the SR-SFIM-based cost functions, such as inconstant levels of process noise and the validity of the EKF's linearization approach\citep{cachim2020}. %

\subsection{Sensor-reduced system}
\label{subsec:redsyssimul}

\begin{figure}[tb]
  \centering
  \includegraphics[width=.8\linewidth]{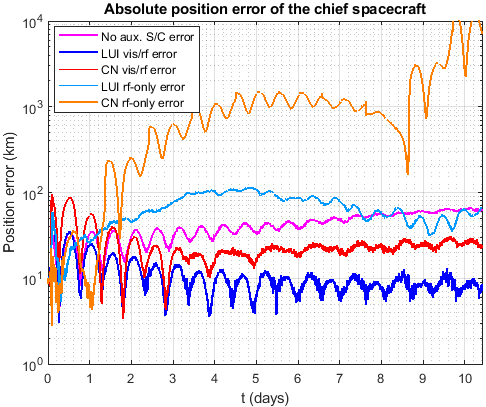}
  \caption{\centering Evolution of the absolute position errors for each optimized configuration of the new SC with the sensor-reduced system.}
  \label{fig:divergenceabs}
\end{figure}

\begin{figure}[tb]
  \centering
  \includegraphics[width=.8\linewidth]{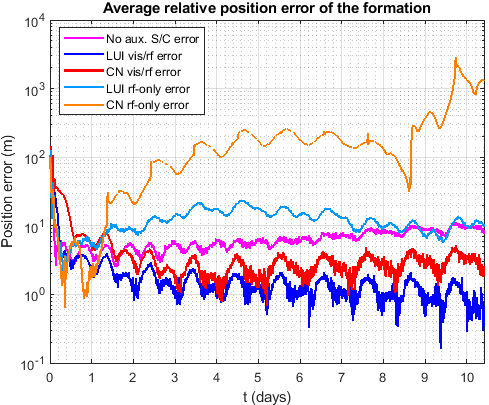}
  \caption{\centering Evolution of the mean formation relative position errors for each optimized configuration of the new SC with the sensor-reduced system.}
  \label{fig:divergencerel}
\end{figure}

To further test the hypothesis described in Section \ref{sec:systsensred}, the sensor-reduced system is also evaluated in a filtering simulation environment with the optimized new spacecraft configurations. With the primary objective being the validation of the sensor-reduced system's error convergence/observability, we first run one simulation with each new spacecraft configuration over a longer duration of 10 orbital periods of the original formation. By removing the relative bearing measurements between the spacecraft in the original formation, but keeping them for the measurements between the chief spacecraft and the new spacecraft with schedule 2, the results in Figures \ref{fig:divergenceabs} and \ref{fig:divergencerel} are produced. To help better gauge whether the filter converges or diverges, the initial absolute position error is increased to $\SI{10}{\km}$ in a random direction.

The plots in Figures \ref{fig:divergenceabs} and \ref{fig:divergencerel} show the absolute and mean relative position error for each of the optimized configurations of the new spacecraft compared to the results obtained when no new spacecraft is present, with the original formation performing ranging-only measurements. Since this scenario is known to be unobservable, it provides a point of reference for the expected divergence rate of the ranging-only estimation. Both RF/Vision configurations of the new spacecraft help reduce the positioning error. While the LUI-optimized RF/Vision-based configuration appears to achieve steady state absolute and relative position errors of $\sim \SI{10}{\km}$ and $\sim \SI{1}{\m}$, respectively, the CN-optimized configuration shows some slow divergence within the considered simulation period. Both RF-Only optimized configurations worsen the positioning performance within the considered time-frame, without reaching a clear steady state.

In \citep{cachim2020}, the different configuration of the process noise covariance matrix $Q$ for each of the optimized configurations may have led to the divergent behaviour that became visible after 5 orbital periods. These results, however, appear to show that, at least for the LUI-optimized RF/Vision-based configuration of the new spacecraft, the system may have enough observability to achieve convergence. Longer simulation periods would help to validate this conclusion.

\section{Conclusions}
\label{sec:concl}

In this paper, the positioning performance of the relative positioning method for the SunRISE mission described in \citep{stuart2017formation} was compared with a modified version that attempts to estimate the absolute states of the spacecraft.
The inclusion of the absolute states of the chief and keplerian dynamics into the navigation solution improved the relative positioning accuracy of the navigation solution from RMS $\SI{3.6}{\m}$ to $\SI{14}{\cm}$. The filter also presents good robustness to poor initial knowledge on the absolute position of the chief spacecraft. The absolute positioning accuracy, however, is considerably worse than that of the GPS-based solution discussed in \citep{stuart2017formation} (by a factor of $\sim 4000$).

An optimization study was performed to find the orbital configuration of a new spacecraft that maximizes the near-GEO formation's positioning accuracy.
The results derived from the continuous-time observability matrix-based optimization are in line with the observations made in \cite{ou2016autonomous, ou2018observability, psiaki1999autonomous}, showing that greater differences in magnitude of gravity acceleration between spacecraft are the main driver in optimizing local observability, as well as wider cross-track motion. 
The addition of both the optimized RF/Vision and the RF-only LUI-optimized new spacecraft configurations shows some tendency for marginal improvements to the absolute positioning performance with schedule 2. However, this tendency is not significant when considering the dispersion of the Monte-Carlo results.
The choice of objective function for the optimization problem may have presented some limitations, such as:
\begin{enumerate*}
	\item the approximation of the original formation to the chief spacecraft;
	\item not accounting for the measurement schedule;
	\item the choice of the SFIM, which more accurately describes the performance of a nonlinear WLS filter than that of the implemented EKF.
\end{enumerate*}

Finally, the observability and performance of the formation when deprived of part of its relative bearing measurements was analysed. The observability analysis showed that the sensor-reduced system should be observable, albeit ill-conditioned. The simulation results indicated that the system shows an apparent divergence of the EKF for most of the considered optimized configurations, with the exception of the LUI-optimized RF/Vision-based system. Testing whether the performance of this system would improve under a square-root filter better suited for ill-conditioned problems would be worth further investigation.

\section*{Acknowledgements}

This work was partially supported by project UIDB/50009/2020 (LARSyS - FCT Plurianual funding 2020-2023), and also by project 24534 - INFANTE (funded by the COMPETE 2020 and Lisboa 2020 programs, under the PORTUGAL 2020 Partnership Agreement, through the European Regional Development Fund).
\bibliographystyle{elsarticle-num}

\bibliography{mybibfile}

\appendix

\section{Simulation Parameters}
\label{app:EKF}

This Appendix lists parameters used in the filter simulations throughout this work.

\apxsub{A.1}{Spacecraft Initial States}{appsubsec:EKFinit}

The initial states of the spacecraft in the formation are described in Table \ref{tab:initstates} with respect to a chief orbit, for which the COE are listed in Table \ref{tab:chieforbit}. These initial states correspond to an example configuration for the SunRISE mission provided in \citep{stuart2017formation} and \citep{hernandez2017satellite}.

\begin{table}[htp]
\centering
\begin{tabular}{lcccccc}
\hhline{=======}
\textbf{COE}                                                    & $\bm{a (km)}$ & $\bm{e}$ &   $\bm{i (deg)}$ &   $\bm{\Omega (deg)}$ &  $\bm{\omega (deg)}$ &  $\bm{\nu (deg)}$ \\ \hline
\textbf{\begin{tabular}[c]{@{}l@{}}Chief \\ Orbit\end{tabular}} & 43399           & 0           & 0           & 0               & 0               & 0            \\ \hhline{=======}
\end{tabular}
\caption{\centering Chief orbit initial COE (Classical Orbital Elements). \label{tab:chieforbit}}
\end{table}

\begin{table*}[htp]
\centering
\begin{tabular}{lcccccc}
\hhline{=======}
\textbf{\# SC} & $\bm{\delta x (km)}$ & $\bm{\delta y (km)}$ & $\bm{\delta z (km)}$ & $\bm{\delta \dot{x} (m/s)}$ & $\bm{\delta \dot{y} (m/s)}$ & $\bm{\delta \dot{z} (m/s)}$ \\ \hline
\textbf{1} & $1.633$  & $4.155$  & $-2.165$ & $-0.1289$ & $-0.1140$ & $8.729\mathrm{e}{-2}$  \\
\textbf{2} & $3.266$  & $0$       & $-1$      & $0$       & $-0.2281$ & $-0.1210$ \\
\textbf{3} & $1.633$  & $-3.655$ & $1.5$     & $9.394\mathrm{e}{-2}$  & $-0.1140$ & $0.1814$  \\
\textbf{4} & $-2.041$ & $-3.443$ & $-3$      & $3.887\mathrm{e}{-2}$  & $0.1425$  & $0$       \\
\textbf{5} & $-3.266$ & $0$       & $1$       & $0$       & $0.2281$  & $-0.1210$ \\
\textbf{6}   & $-2.041$     & $4.443$      & $-1.5$        & $-0.1087$      & $0.1425$       & $0.1814$       \\ \hhline{=======}
\end{tabular}
\caption{\centering Variations in the orbital initial states with respect to the chief orbit for each spacecraft in the formation. \label{tab:initstates}}
\end{table*}


\apxsub{A.2}{New SC process noise configuration}{appsubsubsec:EKFQconfig}

Table \ref{tab:ekfpnparam} shows the chosen $\alpha_{new}$ parameters for each of the configurations with an optimized auxiliary spacecraft shown in the results in Sections \ref{subsec:simulres} and \ref{subsec:redsyssimul}.

\begin{table}[htp]
\centering
\begin{tabular}{llc}
\hhline{===}
\multicolumn{2}{l}{\textbf{\begin{tabular}[c]{@{}l@{}}New SC\\ configuration\end{tabular}}} & $\bm{\alpha_{new}}$ \\ \hline
\multirow{2}{*}{\textbf{RF/Vision}} & \textbf{LUI} & $1\mathrm{e}{4}$ \\
                                    & \textbf{CN}  & $1\mathrm{e}{4}$ \\
\multirow{2}{*}{\textbf{RF-Only}}   & \textbf{LUI} & $1\mathrm{e}{7}$ \\
                                    & \textbf{CN}  & $1\mathrm{e}{8}$ \\ \hhline{===}
\end{tabular}
\caption{\centering EKF’s process covariance matrix $\alpha_{new}$ parameters. \label{tab:ekfpnparam}}
\end{table}

\section{Continuous-time Observability Matrix}
\label{app:obsmat}
Let us consider the following observed dynamic system:
\begin{subequations}
    \begin{align}
        &\Dot{x}(t) = f(x(t)) \label{eq:nonlinsysprop}\\
        &y(t) = h(x(t))\label{eq:nonlinsysobs}
    \end{align}\label{eq:nonlinsys}%
\end{subequations}
with state vector $x$ of size $n$ and observation vector $y$ of size $p$.
The continuous-time observability matrix can be described as the Jacobian of the observation equations and their derivatives with respect to the states, as shown in Eq. \eqref{eq:obsmatcont2} below, in which $m$ is the highest considered differentiation order, and $\ \ \overset{\mathclap{\scriptscriptstyle(m-1)}}{y} \quad (t)$ is the $(\text{m-1})$-th order time-derivative of $y$.

\begin{equation}
    \mathcal{O} = \frac{d}{dx}\begin{bmatrix}
    y(t) \\
    \Dot{y}(t) \\
    \vdots \\
    \quad \overset{\mathclap{\scriptscriptstyle(m-1)}}{y} \quad (t)
    \end{bmatrix} \in \mathbb{R}^{mp \times n}
    \label{eq:obsmatcont2}
\end{equation}

If this matrix is invertible (and therefore has full rank), then the state $x(t)$ can be recovered from the set of observations $y(t)$ and its respective derivatives. The closer to singular the matrix is (i.e, the larger its condition number), the less locally observable the system will be~\citep{krener2009measures}.

For nonlinear systems, this matrix is calculated through the Lie derivatives of the observation equations. A Lie derivative of a function $h$ by a function $f$ is defined as
\begin{equation}
    \mathcal{L}_f(h)(x) = \dfrac{\partial h(x)}{\partial x}f(x) \in \mathbb{R}^{p \times 1}.
\end{equation}

The same Lie derivative of order $k$ is defined as 
\begin{equation}
    \mathcal{L}^k_f(h)(x) = \dfrac{\partial \mathcal{L}^{k-1}_f(h)(x)}{\partial x}f(x) \in \mathbb{R}^{p \times 1}.
\end{equation}

The observability matrix $\mathcal{O}$ of a nonlinear dynamic system such as the one described in \eqref{eq:nonlinsysprop} and \eqref{eq:nonlinsysobs} is therefore defined by these Lie derivatives according to~\cite{krener2009measures}:
\begin{equation}
    \mathcal{O} = \begin{bmatrix}
        \dfrac{\partial h(x)}{\partial x} \\
        \vdots\\
        \dfrac{\partial \mathcal{L}^{m-1}_f(h)(x)}{\partial x}
    \end{bmatrix} \in \mathbb{R}^{mp \times n}.\label{eq:obsmatanalyt}
\end{equation}

For the rank condition to be met, it is necessary that the highest differentiation order under consideration $m$ be greater than $n/p$, otherwise the observability matrix will always be rank defficient. Let us consider a nonlinear system describing the motion of two spacecraft in orbit with two-body problem dynamics and access to relative position measurements:
\begin{equation}
    \begin{cases}
        \begin{bmatrix}
        \Dot{r}_0 \\
        \Dot{v}_0 \\
        \Dot{\delta r}_{1/0} \\
        \Dot{\delta v}_{1/0}
        \end{bmatrix} &= \begin{bmatrix}
        v_0 \\
        -\mu\dfrac{r_0}{\norm{r_0}^3} \\
        \delta v_{1/0} \\
        -\mu\left(\dfrac{r_0 + \delta r_{1/0}}{\norm*{r_0 + \delta r_{1/0}}^3} - \dfrac{r_0}{\norm{r_0}^3}\right)
        \end{bmatrix} \\
        y &= \delta r_{1/0}
    \end{cases}\label{eq:obsmatrelpos}
\end{equation}
where $r_0$ and $v_0$ describe the position and velocity vectors of the chief spacecraft in an inertial frame, $\delta r_{1/0}$ and $\delta v_{1/0}$ the relative position and velocity vectors of the deputy spacecraft in that inertial frame and $\mu$ is the constant gravitational parameter of the central body. The respective local observability matrix requires at least $m=3$ in order to be invertible. This matrix can therefore be defined as:
\begin{equation}
    \mathcal{O}  = \begin{bmatrix}
    0 & 0 & I & 0 \\
    0 & 0 & 0 & I \\
    G_1 - G_0 & 0 & G_1 & 0 \\
    \Dot{G}_1 - \Dot{G}_0 & G_1 - G_0 & \Dot{G}_1 & G_1
    \end{bmatrix}\label{eq:obsmatrelstat}
\end{equation}
where 
\begin{subequations}
    \begin{align}
     G_i &= \dfrac{\mu}{\norm{r_i}^3}\left(3\hat{r}_i\hat{r}_i^T - I\right) \label{eq:Gi}\\
     \Dot{G}_i &= \dfrac{3\mu}{\norm{r_i}^4}\left[v_i\hat{r}_i^T + \hat{r}_iv_i^T - \left(\hat{r}_i^Tv_i\right)\left(5\hat{r}_i\hat{r}_i^T - I\right)\right] \label{eq:dotGi}\\
     \hat{r}_i &= \dfrac{r_i}{\norm{r_i}}
    \end{align}
\end{subequations}
and recalling that $r_1 = r_0 + \delta r_{1/0}$ and $v_1 = v_0 + \delta v_{1/0}$. From the shape of the observability matrix, it is possible to infer that it will have full rank as long as $G_1 - G_0$ has rank 3~\citep{ou2018observability}. 

\pagebreak

\section*{Biography}

\begin{wrapfigure}[7]{l}{25mm} 
	\vspace{-\baselineskip}
    \includegraphics[width=25mm,keepaspectratio]{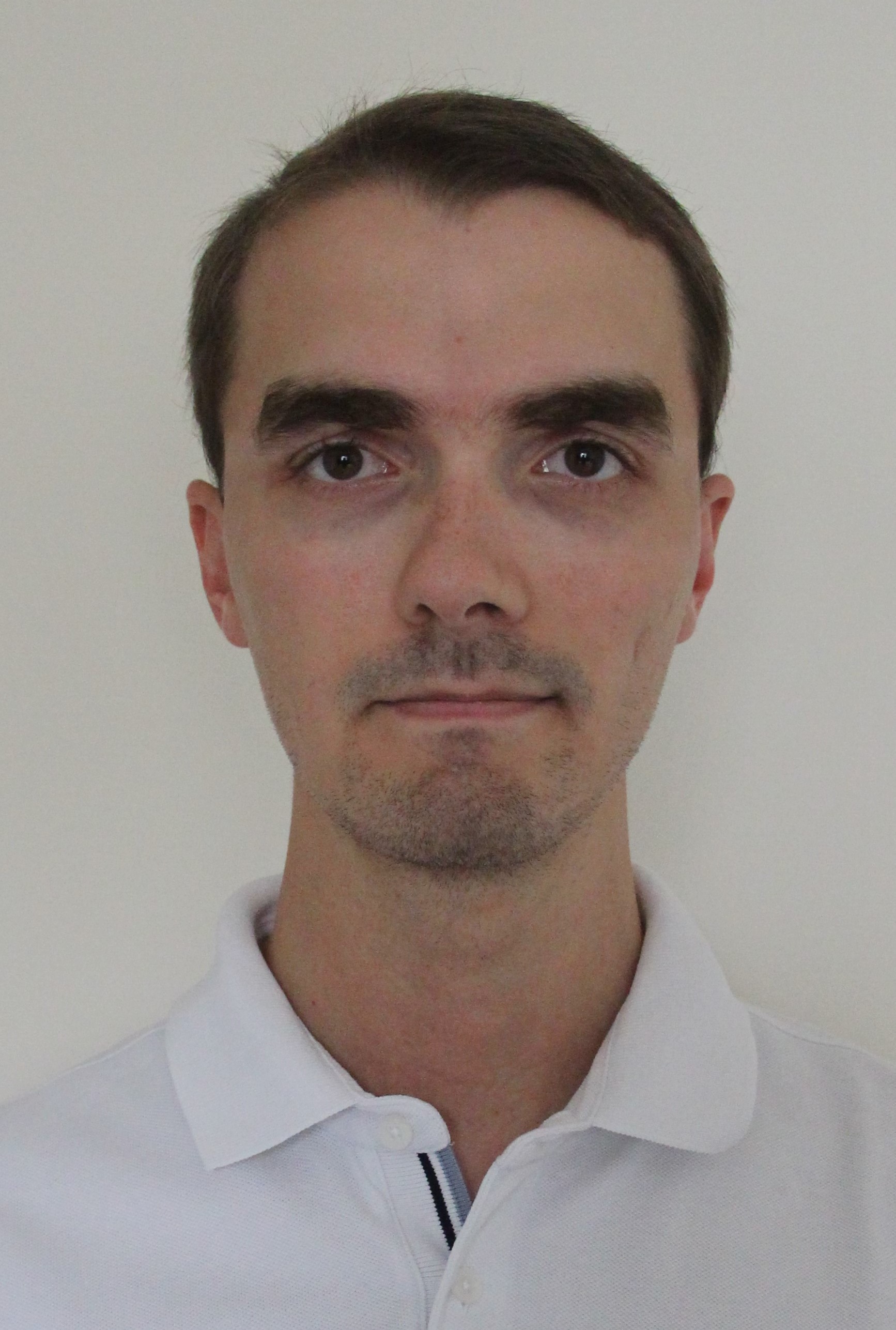}
\end{wrapfigure}\par

\textbf{Pedro Rocha Cachim} received his Masters degree in Aerospace Engineering in 2020 from Instituto Superior Técnico and ISAE-SUPAERO in a double degree exchange program. His studies have been focused in control and navigation systems, primarily towards the aeronautical/space industry. He currently works as a GNC (Guidance, Navigation and Control) Engineer at GMV.\par 

\vspace{10mm}

\begin{wrapfigure}{l}{25mm} 
	\vspace{-\baselineskip}
    \includegraphics[width=25mm,keepaspectratio]{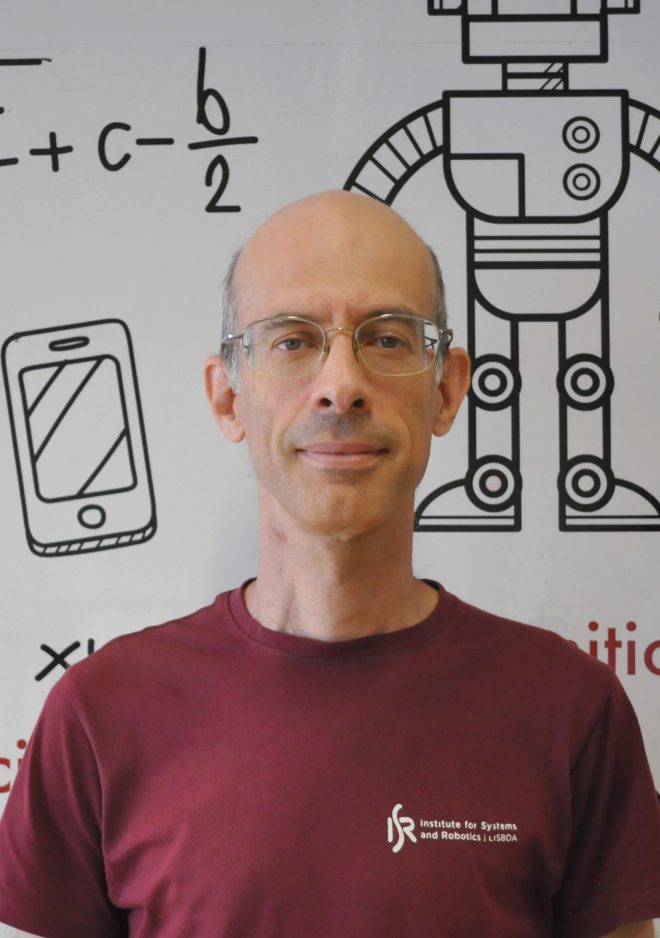}
\end{wrapfigure}\par 

\textbf{João Gomes} received the Diploma, M.Sc. and Ph.D. degrees in electrical and computer engineering from Instituto Superior Técnico (IST) in Lisbon. He is presently an Associate Professor at the Department of Electrical and Computer Engineering of IST, as well as a researcher in the Signal and Image Processing Group of the Institute for Systems and Robotics, in Lisbon. His research interests include localization algorithms for networked systems and GPS-denied environments; channel identification and equalization algorithms in wireless radio and underwater communications; fast algorithms for adaptive filtering; sensor networks.\par 
\vspace{5mm}

\begin{wrapfigure}{l}{25mm} 
	\vspace{-\baselineskip}
    \includegraphics[width=25mm,keepaspectratio]{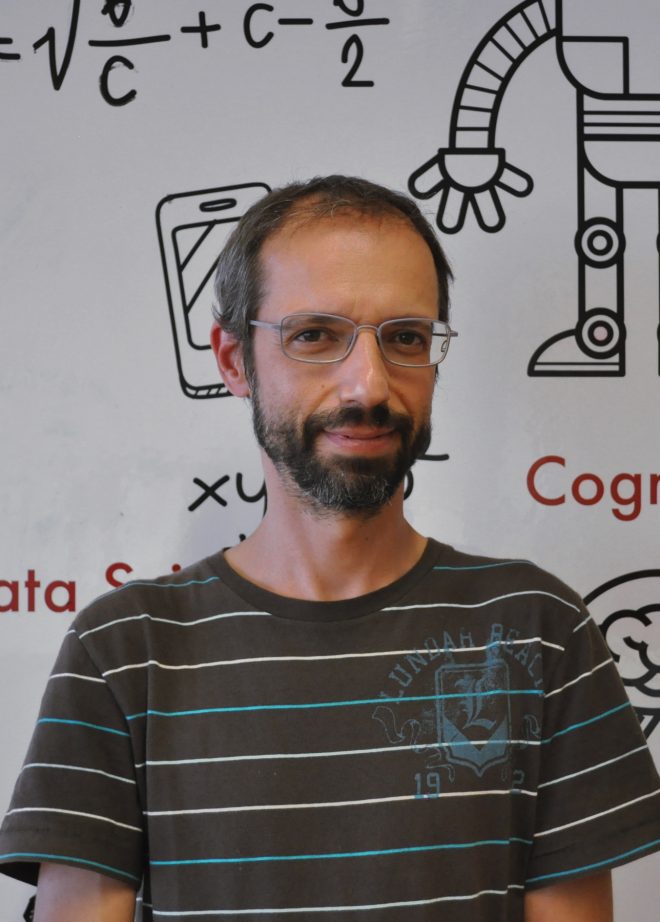}
\end{wrapfigure}\par
\textbf{Rodrigo Ventura} (PhD) is a tenured Assistant Professor of the
Electrical and Computer Engineering Department of Instituto Superior
Técnico (IST), University of Lisbon, and a senior researcher of the
Institute for Systems and Robotics (ISR-Lisbon). He has published more than 130 publications in peer-reviewed international journals and conferences, and is also co-inventor of several national and international patents on innovative solutions for robotic systems. Broadly, his research is focused on the intersection between Robotics and Artificial Intelligence. This research is driven by applications in space robotics, urban search and rescue robotics, aerial robots, and social service robots.\par

\end{document}